\documentclass{article}

 \usepackage[preprint]{neurips_2026}
\makeatletter
\renewcommand{\@noticestring}{}
\makeatother

\usepackage[utf8]{inputenc} 
\usepackage[T1]{fontenc}    
\usepackage{hyperref}       
\usepackage{url}            
\usepackage{booktabs}       
\usepackage{amsfonts}       
\usepackage{nicefrac}       
\usepackage{microtype}      
\usepackage{xcolor}         
\usepackage{footmisc}
\usepackage{tikz}
\usetikzlibrary{positioning,arrows.meta,calc}
\title{The Price of Safety: Benign-Case Utility and Token Overhead of Memory-Poisoning Defenses in LLM Agents}

\author{%
  Pritom Bhowmik \\
  \texttt{bhowmikp1@montclair.edu} \\
}

\usepackage{graphicx}
\hypersetup{hidelinks}
\usepackage{booktabs}
\usepackage{xcolor}
\usepackage{amsmath}
\usepackage{amssymb}
\usepackage{longtable}
\usepackage{float}

\begin{document}

\maketitle

\def\thefootnote{}\footnotetext{Code: \url{https://github.com/pritom02bh/memdefense}}\def\thefootnote{\arabic{footnote}}

\begin{abstract}
Memory-poisoning defenses for LLM agents are usually judged by how well they stop attacks, yet almost all the traffic an agent sees is benign. We measure what these defenses cost when no attack is present. Our harness holds the backbone model, memory backend, retrieval, answer prompt, and judge fixed and changes only the defense. Both LLM-based defenses run on the agent's own backbone, and an integrity check verifies every run. We test four defenses on half of LoCoMo: five long conversations (750 turns and 255 questions per run), with each condition run three times. Sanitization and
provenance checking never fired. LLM-based anomaly detection blocked 9 of 2,250 turns (0.4\%). A read-time reranker dropped 2.2\% of the memories it judged, withholding at least one memory from 14.9\% of questions. A blind audit found content-based justification for only 2 of 50 sampled drops. No defense showed a detectable accuracy cost, at 95\% intervals of $\pm$2.4 to $\pm$4.6 points; the reranker changed accuracy by $+0.6$ points (95\% CI $[-2.6, +3.8]$). Anomaly detection and the stacked condition gained 4.7 points, but a blocked turn changes the memories extracted after it, so we do not count this as a benefit. Token overhead was 0\% for sanitization and provenance, 1.5\% for anomaly detection, 3.0\% for the reranker, and 4.5\% with all four stacked. An earlier version of this study ran the reranker condition on a different model than its baseline. A baseline on that same model shows the swap produced 2.9 of the 4.4 points of accuracy cost we then reported. Attack success rate reveals none of this, so we report benign cost alongside it. We release the harness, labels, and per-run outputs.
\end{abstract}

\section{Introduction}
\label{sec:intro}

LLM agents' traffic is rarely adversarial, yet a memory defense
processes all of it: it scans every turn the agent writes and every
memory the agent reads. The benefit is rare, but the cost is always
there. Most agentic security evaluations measure how well defenses
stop attacks in adversarial conversations. They rarely report what
those defenses cost on ordinary, non-adversarial traffic. Without that
figure, engineers cannot tell whether a defense that blocks attacks
but degrades everyday use is worth deploying.

To answer this, we ask three questions about defenses running on
non-adversarial traffic. \textbf{RQ1:} How does each defense affect
the agent's answer accuracy? \textbf{RQ2:} How many extra tokens does
it add, and how often does it quarantine legitimate memory?
\textbf{RQ3:} When several defenses are stacked, do these costs
compound?

We test four defenses: (1) sanitization, (2) provenance checking,
(3) LLM-based anomaly detection, and (4) retrieval-time reranking.
Each condition shares the undefended baseline's backbone model,
memory backend, retrieval process, answer prompt, and frozen judge,
and both LLM-based defenses run on that same backbone. An integrity
check on every run verifies that each call reached its intended model
and that every turn and verdict was recorded. All traffic is
non-adversarial LoCoMo conversation. We use no adversarial input,
because this study is not about how well defenses stop attacks; prior
evaluations [6, 9, 14] already address that. Because temperature-zero
runs are not fully deterministic, we ran each condition three times.

An earlier version of this study ran the reranker's verdicts and its
condition's answers on a different model than the baseline. A
baseline run on that same model shows that the swap accounted for 2.9
of the 4.4 points of accuracy cost we then attributed to the
reranker; at equal model, the reranker's remaining 1.5 points could
not be distinguished from zero.

Sanitization and provenance never fired. Anomaly detection blocked
the same three turns in every replicate: 9 of 2,250 turns (0.4\%). No
defense showed a detectable accuracy cost, at 95\% intervals of
$\pm$2.4 to $\pm$4.6 points (RQ1); the reranker changed accuracy by
$+0.6$ points (95\% CI $[-2.6, +3.8]$). Anomaly detection and the
stacked condition gained 4.7 points. We do not count this as a
benefit, because a blocked turn changes the memories extracted after
it, and even provenance, which never fired, moved one conversation by
9 points. Token overhead ranged from 0\% for sanitization and
provenance to 1.5\% for anomaly detection, 3.0\% for the reranker,
and 4.5\% when stacked. The reranker dropped 2.2\% of the memories it
judged, withholding at least one memory from 14.9\% of questions, and
a blind audit found 48 of 50 sampled drops unjustified (RQ2). Stacked
accuracy matched anomaly detection's in every conversation, and its
token overhead was the sum of its two LLM stages (RQ3). The reranker
fired more than five times as often as the anomaly detector
(2.2\% vs.\ 0.4\%). Stage and implementation vary together in our
design, so we report this as an association.

We make four contributions: (1)~a replicated harness, released with
the paper, that isolates the benign-case cost of four memory defenses
on a shared backbone with per-run integrity checks; (2)~to our
knowledge, the first comparison of write- and read-time defenses
under identical non-adversarial conditions; (3)~a same-model baseline
showing that a model swap produced most of a previously reported
defense cost; and (4)~evidence that accuracy alone misses how often a
defense withholds memory.


\section{Related Work}
\label{sec:related}

\subsection{Memory Poisoning in LLM Agents}
\label{sec:memory-poisoning}
An agent uses persistent memory to keep information across sessions.
That memory also gives an attacker a foothold that outlasts any single
interaction. Content planted in an agent's memory in one session can
resurface in later sessions, and the agent has no reason to suspect it
[1, 2]. Retrieval ranks memories by similarity and never checks trust.
A convincing piece of adversarial content can therefore rank as highly
as a legitimate memory [3, 4]. Recent research maps attacks across the
write, retrieval, and action stages [5, 6]. Other studies locate the
trust boundary, where untrusted text enters the agent's record [7, 8].

\subsection{Proposed Defenses}
\label{sec:related-defenses}
Defenses are usually grouped by the stage at which they act.
Write-time defenses act before the agent stores content in its memory.
Some scan incoming content [9]. Others tag each write with provenance
and admit only verified entries [11]. A third kind scores each
candidate entry for anomalies before storing it, for example by
embedding similarity [12]. Read-time defenses leave storage unchanged
and act on what retrieval returns. Some weigh retrieved memories by
trust [9], others filter suspicious retrieved documents [10], and some
answer from each passage separately and then aggregate the answers
[13]. Act-time defenses check a memory just before it affects a
decision [14]. Some designs place checks at more than one stage
instead of relying on a single one [15].

\subsection{The Attack-Success-Rate Gap}
\label{sec:related-gap}
Most memory-defense papers build an attack, test whether the defense
stops it, and report the attack success rate [9, 14]. A few add a
single utility figure. A-MemGuard reports a minimal utility cost [14],
and ASB combines utility and security into one metric [6]. For
prompt-injection defenses, AgentDojo reports utility without attacks
[23]. There, a detector's false positives cut GPT-4o's benign utility
from 69.0\% to 41.5\%. None of these reports a memory defense's benign
cost as accuracy change, token overhead, and false-quarantine rate
together, replicated across runs. Other studies, from the systems
side, benchmark memory architectures on accuracy, latency, and token
cost [16, 17]. These benchmarks usually test the memory system alone,
with no defense added [18, 19]. So security work studies whether
defenses stop attacks, and systems work studies what memory itself
costs. Neither measures what a memory defense costs on non-adversarial
traffic, and this paper fills that gap.


\section{Threat Model and Defenses}
\label{sec:threat}

\subsection{Pipeline}
\label{sec:threat-pipeline}

The agent we analyze follows the usual structure of memory-augmented systems. A conversation arrives one turn at a time. At each turn, an extraction step decides what, if anything, to remember and writes it to a vector store. For each question, the system retrieves the most similar memories and adds them to the model's context. Memory extraction, answering, and both LLM-based defenses all call the same backbone model. 

A defense can intervene at three points. A write-time defense screens content before storing it. A read-time defense filters memories as they enter the context. An act-time defense checks a memory before the agent uses it to take an action. Most proposed defenses target one of these points. We ask whether the intervention point, rather than the mechanism, predicts benign cost. Figure~\ref{fig:pipeline} shows our instantiation. 

The threat is an attacker who inserts text that becomes a memory and is later retrieved as fact [5, 7]. We do not test the attack. We measure what the defenses cost when no injection occurs.

\begin{figure}[H]
\centering
\footnotesize
\begin{tikzpicture}[
  node distance=20mm and 5.5mm,
  box/.style={draw=black!60, rounded corners=1.5pt, align=center,
    text width=25mm, minimum height=14mm, inner sep=1.5pt},
  wgate/.style={box, text width=31mm, fill=blue!7,
    draw=blue!60!black, thick},
  rgate/.style={box, text width=31mm, fill=red!7,
    draw=red!60!black, thick},
  meas/.style={box, dashed, fill=black!4},
  flow/.style={-{Stealth[length=1.8mm]}, semithick, black!75},
  drop/.style={-{Stealth[length=1.5mm]}, dashed, black!60},
  note/.style={font=\scriptsize, inner sep=1pt},
  lane/.style={font=\scriptsize\itshape, text=black!60,
    inner sep=0pt},
]
\node[box] (turn) {Conversation turn\\(150 per conv.)};
\node[wgate, right=of turn] (wg) {\textbf{Write-time defenses}\\
  Sanitize (regex)\\
  Provenance (origin)\\
  Anomaly (backbone LLM)};
\node[box, right=of wg] (ext) {Memory extraction\\
  (backbone LLM,\\via Mem0)};
\node[box, right=of ext] (store) {Vector store\\
  (namespace per cell)};

\node[box, below=of store] (ret)
  {Question $\to$\\retrieve top-10\\by similarity};
\node[rgate, left=of ret] (rr) {\textbf{Read-time defense}\\
  Rerank (backbone LLM):\\
  trust or drop each\\
  memory, judged alone};
\node[box, left=of rr] (ans) {Answer from kept\\
  memories\\(backbone LLM)};
\node[meas, left=of ans] (judge)
  {LLM judge (other\\provider) or\\abstention detector};

\node[meas, text width=40mm, minimum height=5mm, fill=white]
  at ($(wg.south)!0.5!(rr.north)$) (qlog) {Quarantine log};
\node[note, right=1.5mm of qlog] {$\to$ FQR, Eq.~\eqref{eq:fqr}};

\draw[flow] (turn) -- (wg);
\draw[flow] (wg) -- node[note, above]{kept} (ext);
\draw[flow] (ext) -- (store);
\draw[flow] (store) -- (ret);
\draw[flow] (ret) -- (rr);
\draw[flow] (rr) -- node[note, above]{kept} (ans);
\draw[flow] (ans) -- (judge);
\draw[drop] (wg.south) -- node[note, right]{blocked}
  (qlog.north -| wg.south);
\draw[drop] (rr.north) -- node[note, right]{dropped}
  (qlog.south -| rr.north);

\node[note, below=1mm of judge.south west, anchor=north west]
  {$\to$ paired $\Delta_d$ vs.\ baseline, Eq.~\eqref{eq:bud}};
\node[lane, above=1mm of turn.north west, anchor=south west]
  {Write path: once per turn};
\node[lane, below=1mm of ret.south east, anchor=north east]
  {Read path: once per question};

\path let \p1 = ($(ret.east)-(judge.west)$) in
  node[meas, text width=\x1-3pt, minimum height=6mm,
    anchor=north west]
  (proxy) at ($(judge.south west)+(0,-7mm)$)
  {Every LLM call passes a logging proxy tagged \emph{base},
   \emph{defense}, or \emph{judge} $\to$ token overhead $\tau_d$,
   Eq.~\eqref{eq:tau}, and the per-run integrity check};
\end{tikzpicture}
\caption{Measurement pipeline. The write path runs once per
conversation turn. The read path runs once per question, so the
reranker judges the same memory repeatedly and without its source
turn. Extraction, answering, anomaly detection, and the reranker all
call the same backbone model. Each condition enables one defense, and
the stacked condition enables all four. Everything else is held
fixed.}
\label{fig:pipeline}
\end{figure}

\subsection{Write-Time Defenses}
\label{sec:threat-write}
Three of our four defenses check each turn before extracting memories, so blocked turns never reach the memory layer. The defenses work differently and use different amounts of tokens, but they all run at the same point. Comparing them with the reranker helps show whether the cost to benign users depends on where a defense runs, while token cost depends on how it works. 

Sanitization checks each turn against eight regular expressions for injection phrases, such as ``ignore previous instructions'' or ``system prompt.'' It blocks any match without calling a model. Provenance blocks turns that did not originate from the user. LoCoMo turns have no origin field and default to the user, so provenance never triggers here. Its $\Delta_d$ therefore measures run-to-run noise. 

Anomaly detection is the only write-time defense that calls a model. It sends each turn by itself, in the form ``speaker: text,'' to the backbone at temperature 0. The turn is blocked if the reply is \texttt{SUSPICIOUS} rather than \texttt{BENIGN}. Failed calls are retried with backoff. If a call still fails, the run stops, ensuring no turn passes unscreened.

\subsection{Read-Time Interception}
\label{sec:threat-read}
The reranker leaves storage unchanged and works when a question is answered. Like the baseline, it retrieves the 10 most similar memories. It then sends each memory to the backbone separately, at temperature 0. The call includes neither the question nor the memory's source turn. The prompt warns that some memories may be untrustworthy or planted and asks for \texttt{TRUST} or \texttt{DROP}. Dropped memories are left out of the context, and the answer prompt remains the same as the baseline. Failed calls are handled as they are at write time, so no memory is kept by default.

Two differences from write-time screening shape our results. First, there is frequency. A write-time filter judges each turn once, while the reranker judges a memory every time it is retrieved. So, a memory it distrusts can be withheld again for later questions. Second, there is evidence. At write time, a defense sees the whole turn and its speaker. At read time, it sees only a short, extracted string, without context. Anomaly detection and the reranker use the same backbone but different prompts and inputs. So, our design cannot fully separate the stage from the implementation (Section~\ref{sec:limitations}).

\subsection{Stacked Composition}
\label{sec:threat-stacked}
The stacked setup applies all four defenses to the store. This follows the common advice to use layers of protection, since no single defense can cover every possible attack. Each added filter may also block valid input, but whether those effects build up is something we need to test. Because the defenses were wrapped around one another, the reranker could no longer access the store and failed on every question. We fixed this before the reported runs. The bug caused a crash, but a silent failure could have appeared to be a valid result (Appendix~\ref{app:pitfalls}).


\section{Measurement Harness}
\label{sec:harness}

\subsection{Isolation}
\label{sec:measurement-isolation}
We measure the marginal cost of each defense while holding everything else fixed (Figure~\ref{fig:pipeline}). Across conditions, the backbone model, memory backend, embedding model, top 10 retrieval, answer prompt, question set, and judge stay exactly the same. Only the defense changes. Each cell, one run of one condition on one conversation, writes to its own namespace. Nothing one cell stores or blocks can affect another.

\subsection{Integrity Check}
\label{sec:integrity}
An integrity check reads each cell's proxy log and verifies three
things. First, every call went to the model assigned to its step, and
a live check would have stopped any run that broke this rule. Second,
each of the 150 turns went through extraction, and none of them picked
up leftover messages from an earlier, interrupted attempt. Third, each
defense gave exactly one verdict for every turn it screened and every
memory it judged, and the agent answered every question exactly once.
All 135 cells pass this check: the 90 main cells and the 45 cells of
the model-swap baseline (Section~\ref{sec:protocol}). Two of them pass
with a known exception, which Appendix~\ref{app:pitfalls} explains.

\subsection{Per-Purpose Cost Attribution}
\label{sec:measurement-cost}
Every model call passes through a local logging proxy. Each call is tagged with its purpose (\emph{base} for memory extraction and answering, \emph{defense}, or \emph{judge}) and with the cell and step it belongs to. Summing tokens by tag separates the defense's cost from the agent's own cost in each cell, so the token overhead $\tau_d$ (Section~\ref{sec:metrics}) is a direct count. Base and defense calls run on the same model, so a single tokenizer counts both.

\subsection{Dataset and Evidence Alignment}
\label{sec:dataset}
We evaluate on LoCoMo [20], which contains ten conversations pairing long, multi-session dialogues with questions and answers. To keep 90 cells affordable, we use the first five conversations in release order, avoiding selection based on results. We truncate each conversation to its first 150 turns. 

LoCoMo's questions were written for full conversations of 369 to 680 turns. At 150 turns, only 266 of the 999 questions (27\%) keep all their evidence turns. The rest cannot be answered by any agent, defended or not, so keeping them would measure truncation rather than defense. We keep only the 266 and cap each conversation at 75 questions, trimming only conv-26 from 86 to 75. This leaves 255 questions per replicate: 200 core and 55 adversarial (Appendix~\ref{app:retention}). We fixed this filter before any run and applied it identically across conditions. 

Over three replicates, each condition yields 765 graded answers. Each write-time defense makes 2,250 verdicts, and the reranker makes 7,650. Five conversations limit how far our results generalize, so Section~\ref{sec:metrics} also tests with the conversation as the unit.

\subsection{Judge and Calibration}
\label{sec:judge}

\paragraph{Core answers} An LLM judge from a different provider than the backbone grades the core answers, so no model grades its own output. Before the experiment, we hand-graded 24 core answers, six per category, from a separate 60-turn run on conv-26. The judge agreed with us on 23 of them ($\kappa = 0.909$). With only 24 items, the exact 95\% interval for this agreement is wide, from 79 to 99.9\%. The paired design limits the risk: every condition uses the same frozen judge and prompt, so errors unrelated to the condition cancel in $\Delta_d$. 

\paragraph{Adversarial answers} Adversarial questions ask about things the conversation does not support, so we score whether the agent abstains. A regex that matches refusal phrases gives the first score. An external annotator, blind to condition, then labeled 52 of the 55 answers the regex scored as failures and a sample of 56 it scored as abstentions. A label covers every answer with the same text, so these 108 answers contain only 23 distinct texts. 

The regex agreed with the annotator on 91 of the 108 answers ($\kappa = 0.68$) and on 20 of the 23 texts ($\kappa = 0.73$; Table~\ref{tab:calibration}). Every error was a refusal that the regex scored as a failure. The final score uses the hand label where one is available and the regex otherwise. A broader refusal detector agrees with the regex on every unlabeled answer. The three unlabeled failures come from conv-30 cells rebuilt after labeling, and each makes a substantive claim, so they keep the regex's call. These scores do not affect the core $\Delta_d$, the false quarantine rate, or $\tau_d$.

\begin{table}[H]
\caption{Adversarial answers labeled by hand, blind to condition. Rows
group answers by the regex's call. A label covers every answer with
the same text, so we count both answers and distinct texts.}
\label{tab:calibration}
\centering
\small
\setlength{\tabcolsep}{6pt}
\renewcommand{\arraystretch}{1.15}
\begin{tabular}{@{}l rr rr@{}}
\toprule
 & \multicolumn{2}{c}{\textbf{Answers}}
 & \multicolumn{2}{c}{\textbf{Distinct texts}} \\
\cmidrule(lr){2-3}\cmidrule(l){4-5}
\textbf{Regex said} & $\boldsymbol{n}$ & \textbf{Agree}
 & $\boldsymbol{n}$ & \textbf{Agree} \\
\midrule
Abstained (sample) & 56  & 56   & 7  & 7  \\
Failed             & 52  & 35   & 16 & 13 \\
\midrule
Total              & 108 & 91   & 23 & 20 \\
Cohen's $\kappa$   &     & 0.68 &    & 0.73 \\
\bottomrule
\end{tabular}
\end{table}


\section{Metrics and Estimation}
\label{sec:metrics}

\subsection{Notation}
Let $\mathcal{C}$ be the five conversations and $\mathcal{R}$ the
three replicates. The conditions are $d \in \mathcal{D} =
\{\varnothing, \mathrm{san}, \mathrm{prov}, \mathrm{anom},
\mathrm{rer}, \mathrm{stk}\}$, where $\varnothing$ is the undefended
baseline. A \emph{cell} is one triple $(d, c, r)$, so the design has
$|\mathcal{D}||\mathcal{C}||\mathcal{R}| = 90$ cells. For
conversation $c$, $Q_c$ is the question set kept in
Section~\ref{sec:dataset}, split into core $Q_c^{\mathrm{core}}$ and
adversarial $Q_c^{\mathrm{adv}}$. We write $y_{d,c,r,q} \in \{0,1\}$
for whether the answer to $q$ in cell $(d, c, r)$ is correct.

\subsection{Collapsing Replicates}
The pipeline is not fully deterministic, even at temperature zero.
Across baseline replicates of the same conversation, core accuracy
changed by up to 3.0 points and the number of stored memories by up to
21. Replicates absorb this noise; they do not enlarge the sample. We
therefore collapse them with a majority vote over
$\mathcal{R}_{d,c,q}$, the replicates that graded $q$. With three
graded replicates, an item counts as correct when at least two are
correct:
\begin{equation}
\hat{y}_{d,c,q} = \mathbf{1}\!\left[\frac{1}{|\mathcal{R}_{d,c,q}|}
\sum_{r \in \mathcal{R}_{d,c,q}} y_{d,c,r,q} > \tfrac{1}{2}\right]
\label{eq:majority}
\end{equation}

\subsection{Benign Utility Delta}
Baseline accuracy ranges from 48.5\% to 80.0\% across conversations
(SD 11.5 points). That spread is more than twice the largest effect
we measure, so an unpaired comparison would bury effects of a few
points. Every condition answers the same questions, so we compare
within each question. Let $\tilde{Q}_{d,c}$ be the questions graded
under both $d$ and the baseline:
\begin{equation}
\Delta_{d,c} = \frac{1}{|\tilde{Q}_{d,c}|}
\sum_{q \in \tilde{Q}_{d,c}}
\left(\hat{y}_{d,c,q} - \hat{y}_{\varnothing,c,q}\right)
\label{eq:convdelta}
\end{equation}
The benign utility delta weights conversations equally, so a long
conversation cannot dominate:
\begin{equation}
\Delta_d = \frac{1}{|\mathcal{C}|} \sum_{c \in \mathcal{C}}
\Delta_{d,c}
\label{eq:bud}
\end{equation}
We compute $\Delta_d$ separately for core and adversarial questions.
On adversarial questions, a correct answer is an abstention. We also
report $\Delta_d$ within each replicate, without the vote. The
model-swap comparisons of Section~\ref{sec:protocol} use the same
paired $\Delta$, with other cells in place of $d$ and the baseline.

\subsection{Token Overhead}
Let $T^{\mathrm{def}}_d$ be the tokens tagged \emph{defense} across
the 15 cells of condition $d$, and $T^{\mathrm{base}}_{\varnothing}$
the tokens tagged \emph{base} across the 15 baseline cells:
\begin{equation}
\tau_d = \frac{T^{\mathrm{def}}_d}{T^{\mathrm{base}}_{\varnothing}}
\label{eq:tau}
\end{equation}
We fix the denominator at the baseline's base tokens, so run-to-run
changes in the agent's own spend do not enter $\tau_d$. We report
tokens instead of latency, which provider rate limits inflate, or
dollars, which change with prices.

\subsection{False Quarantine Rate}
A quarantine is a defense's decision to block a turn or drop a
retrieved memory. In each cell, $n^{\mathrm{quar}}_{d,c,r}$ counts
quarantines and $n^{\mathrm{seen}}_{d,c,r}$ counts decisions: turns
screened at write time, or memories judged at read time. We pool over
cells, so every decision counts equally:
\begin{equation}
\mathrm{FQR}_d = \frac{\sum_{c,r} n^{\mathrm{quar}}_{d,c,r}}
{\sum_{c,r} n^{\mathrm{seen}}_{d,c,r}}
\label{eq:fqr}
\end{equation}
For the stacked condition, we report the write-time and read-time
stages separately. No content in our corpus was planted, so every
quarantine is a false alarm. A false alarm is not always harmful,
because the dropped memory may be irrelevant to the question or truly
unreliable. FQR measures how often a defense fires without an attack;
$\Delta_d$ measures what that costs in accuracy. Two companion rates
show the reach of read-time quarantine: the share of questions that
lose at least one memory, and the share of stored memories dropped at
least once. All three rates come straight from the defense's
decisions, so retrieval and generation cannot mask them.

\subsection{Uncertainty}
We report three tests. The headline is a stratified bootstrap
interval: we resample questions within each conversation $B = 10^4$
times, recompute $\Delta_d$ with equal conversation weights, and take
the 2.5th and 97.5th percentiles [21]. This interval reflects
question sampling; run-to-run noise enters only through the majority
vote. The exact McNemar test [22] uses only the discordant pairs,
which number 5 to 26 for core accuracy, too few for the $\chi^2$
approximation. The exact sign-flip test treats the conversation as
the unit. It enumerates all $2^5 = 32$ sign assignments of the
$\Delta_{d,c}$ and reports the share whose mean is at least as far
from zero as $\Delta_d$, so its $p$-value cannot fall below
$2/32 = 0.0625$. Interval half-widths range from 2.4 to 4.6 points.
We therefore read every null as no effect detectable at this
resolution, not as zero cost (Section~\ref{sec:limitations}).


\section{Experimental Setup}
\label{sec:experiment}

\subsection{Models and Configuration}
\label{sec:models}

Table~\ref{tab:models} lists all models. One backbone, \texttt{gemini-3.1-flash-lite}, handles memory extraction, question answering, and both LLM-based defenses. Each LLM defense therefore runs on the same model as the agent it protects, so differences between conditions do not come from a model change. The backbone is a hosted model without a pinned version (Section~\ref{sec:limitations}). Memory runs on Mem0 2.0.15 [16] with a local Qdrant store. Retrieval returns the top 10 memories by embedding similarity, with no threshold. All generation uses temperature 0, and extraction has an explicit 4,000-token output limit. We chose this backbone because 90 cells on a larger Gemini model cost too much. Switching models also exposed a silent extraction failure (Appendix~\ref{app:pitfalls}).

\begin{table}[H]
\caption{Models used in the experiment. The agent performs memory
extraction and answering; the LLM defenses are anomaly detection and
the reranker. The judge's provider differs from the backbone's.
\texttt{gpt-4o-mini} appears only in the model-swap baseline.}
\label{tab:models}
\centering
\small
\setlength{\tabcolsep}{5pt}
\begin{tabular}{@{}l l l@{}}
\toprule
\textbf{Role} & \textbf{Model} & \textbf{Provider} \\
\midrule
Agent                     & \texttt{gemini-3.1-flash-lite}
  & Google \\
LLM defenses              & \texttt{gemini-3.1-flash-lite}
  & Google \\
Embedding                 & \texttt{text-embedding-3-small}
  & OpenAI \\
Judge                     & \texttt{claude-haiku-4-5-20251001}
  & Anthropic \\
\midrule
Model-swap baseline only  & \texttt{gpt-4o-mini}
  & OpenAI \\
\bottomrule
\end{tabular}
\end{table}

\subsection{Protocol}
\label{sec:protocol}
We run six conditions: the baseline, each write-time defense alone, the reranker alone, and all four defenses stacked. Each cell has two phases (Figure~\ref{fig:pipeline}). First, the agent ingests the conversation turn by turn, with the defense active at its stage. It then answers the questions from the resulting store, and the judge or the abstention scorer grades each answer. We run each condition three times on each of the five conversations. 

On conv-42 and conv-43, every cell answers again on the backbone from the stores built for the earlier version of this study. Rerunning their write-time screens on the backbone blocked no turn, so no store holds a turn the corrected pipeline would block. On the other three conversations, the corrected anomaly screen blocked a turn that the old stores contained. We therefore rebuilt every store for these three conversations in one batch. Mem0 writes the current date into its extraction prompt, so we fixed that date for the whole batch. Within each conversation, every condition reads memory built in the same period, so no paired $\Delta$ compares stores built at different times. 

To measure the model confound in our earlier version, we add a model-swap baseline of 45 cells on all five conversations. It reuses the earlier version's 15 baseline cells and 15 reranker cells; the reranker cells ran their verdicts and answers on \texttt{gpt-4o-mini}. Fifteen new cells answer from the same baseline stores with \texttt{gpt-4o-mini} and no defense. Comparing these cells splits the earlier reranker delta into two parts: the effect of the answer model and the effect of the reranker at the same model. 

For every cell, we record the memory count, tokens by purpose, per-question grades, and the quarantine log. Cells replaced by rebuilds stay on disk, and a manifest fixes the one cell used for each condition, conversation, and replicate. Every reported number traces back to the cells in that manifest.


\section{Results}
\label{sec:result}
\begin{figure}[H]
\centering
\includegraphics[width=3.4in]{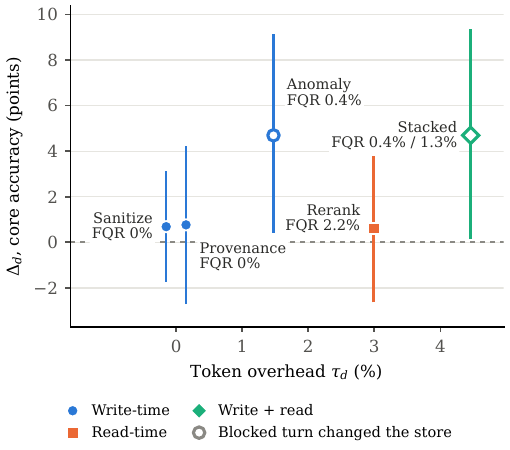}
\caption{Accuracy change against token overhead. Bars are 95\%
bootstrap intervals. Hollow markers show conditions whose write-time
screen blocked turns in three of the five conversations. A blocked
turn changes the memories extracted after it, so their $\Delta_d$
compares different stores. Sanitize and provenance both cost 0\% and
are offset for visibility. Stacked FQR is given for its write-time
and read-time stages.}
\label{fig:costutility}
\end{figure}

Table~\ref{tab:main} gathers the main results.
Figure~\ref{fig:costutility} plots accuracy change against token
cost, and Figure~\ref{fig:modelswap} splits the reranker cost reported
in our earlier version.

\begin{table}[H]
\caption{Benign-case results relative to the undefended baseline.
$\Delta_d$ is the paired change in core accuracy, in points (majority
vote over three replicates, conversations weighted equally), with its
stratified bootstrap 95\% CI and exact McNemar and sign-flip
$p$-values. Per replicate gives $\Delta_d$ within each replicate.
$\tau_d$ is defense tokens over baseline tokens. FQR is quarantines
per decision; for the stacked condition it is given for the
write-time and read-time stages.}
\label{tab:main}
\centering
\small
\setlength{\tabcolsep}{4pt}
\renewcommand{\arraystretch}{1.15}
\begin{tabular}{@{}l l r r@{,\,}l r r r r r@{}}
\toprule
\textbf{Defense} & \textbf{Stage} & $\boldsymbol{\Delta_d}$
  & \multicolumn{2}{c}{\textbf{95\% CI}}
  & $\boldsymbol{p_{\mathrm{McN}}}$
  & $\boldsymbol{p_{\mathrm{flip}}}$ & \textbf{Per replicate}
  & $\boldsymbol{\tau_d}$ & \textbf{FQR} \\
\midrule
Sanitize   & write & $+0.7$ & $[-1.7$ & $+3.1]$ & 1.000 & 1.000
  & $-0.9,\ -0.4,\ +0.8$ & 0.0\% & 0.0\% \\
Provenance & write & $+0.8$ & $[-2.7$ & $+4.2]$ & 1.000 & 1.000
  & $+1.4,\ -0.4,\ +0.7$ & 0.0\% & 0.0\% \\
Anomaly    & write & $+4.7$ & $[+0.4$ & $+9.1]$ & 0.152 & 0.125
  & $+3.6,\ +3.6,\ +4.2$ & 1.5\% & 0.4\% \\
\midrule
Rerank     & read  & $+0.6$ & $[-2.6$ & $+3.8]$ & 0.754 & 1.000
  & $-0.5,\ +1.3,\ +0.1$ & 3.0\% & 2.2\% \\
Stacked    & both  & $+4.7$ & $[+0.1$ & $+9.4]$ & 0.169 & 0.125
  & $+1.7,\ +3.0,\ +4.2$ & 4.5\% & 0.4 / 1.3\% \\
\bottomrule
\end{tabular}
\end{table}

\subsection{Benign Utility (RQ1)}
\label{sec:result-benign}
Sanitization and provenance never fired. Each screened 2,250 turns
and quarantined none, so their $\Delta_d$ reflects run-to-run noise:
$+0.7$ and $+0.8$ points, with both intervals including zero. Within
single replicates, the two range from $-0.9$ to $+1.4$ points.

The reranker shows no detectable cost either. Its $\Delta_d$ is
$+0.6$ points (95\% CI $[-2.6, +3.8]$). Four core questions went from
right to wrong and six went the other way (McNemar $p = 0.75$). Its
per-replicate values are $-0.5$, $+1.3$, and $+0.1$ points.

Anomaly detection and the stacked condition both gained 4.7 points.
Their bootstrap intervals exclude zero, at $[+0.4, +9.1]$ and
$[+0.1, +9.4]$, but neither McNemar ($p = 0.15$ and $0.17$) nor the
sign-flip test ($p = 0.125$) rejects zero. We do not read this gain as
a benefit. The anomaly screen blocked one turn in each of conv-26,
conv-30, and conv-41, the same turn in every replicate. A blocked turn
changes the memories extracted after it. In conv-26, only 31\% of
anomaly detection's answers match the baseline's word for word,
against 100\% for sanitization. The bootstrap reflects question
sampling, not this change of store. Separate stores alone can move a
conversation by several points: in conv-42, sanitization and
provenance, which never fire, moved by $+6.1$ and $+9.1$ points
(Figure~\ref{fig:perconv}).

Adversarial questions show no cost from any defense. The baseline
abstained on 94.7\% to 96.9\% of them per replicate. Pooled changes
range from $-1.2$ to $+3.1$ points, with every McNemar $p$ at 0.50 or
above. Single replicates range from $-3.4$ to $+4.3$ points, and the
reranker changed abstention by $0.0$, $0.0$, and $+1.0$ points.

\subsection{Model Swap in the Earlier Version}
\label{sec:result-swap}
\begin{figure}[H]
\centering
\includegraphics[width=3.4in]{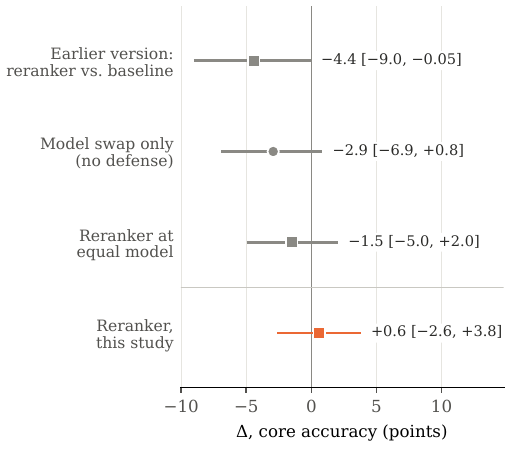}
\caption{The source of the earlier version's reranker cost. The top
three rows use the earlier version's cells, where the reranker's
verdicts and its condition's answers ran on \texttt{gpt-4o-mini}. The
bottom row is this study's reranker on the matched backbone. The model
swap alone ($-2.9$) and the reranker at equal model ($-1.5$) add up to
the earlier $-4.4$. Bars are 95\% bootstrap intervals.}
\label{fig:modelswap}
\end{figure}

Our earlier version reported that the reranker cut core accuracy by
4.4 points (95\% CI $[-9.0, -0.05]$). In those cells, the reranker's
verdicts and its condition's answers ran on \texttt{gpt-4o-mini},
while the baseline answered on the backbone.
Figure~\ref{fig:modelswap} splits that delta using the model-swap
baseline. Swapping only the answer model, with no defense, changed
accuracy by $-2.9$ points $[-6.9, +0.8]$. Adding the reranker at equal
model changed it by a further $-1.5$ points $[-5.0, +2.0]$. Neither
part alone differs detectably from zero (McNemar $p = 0.36$ and
$0.30$). The model swap accounts for about two-thirds of the earlier
cost. On the matched backbone, the reranker's change is $+0.6$
points.

\subsection{Token Overhead (RQ2)}
\label{sec:benign-token}
Token cost follows mechanism. Sanitization and provenance make no
model call, so $\tau_d = 0$. Anomaly detection makes one call per
turn, 750 per replicate, for $\tau_d = 1.5\%$. The reranker makes one
call per retrieved memory, 2,550 per replicate, for 3.0\%. The stacked
condition pays both, 4.5\%, close to their sum.

\subsection{False Quarantine (RQ2)}
\label{sec:benign-quarantine}
Anomaly detection blocked 9 of 2,250 turns (FQR 0.4\%), the same three
turns in every replicate. The reranker dropped 169 of the 7,650
memories it judged (FQR 2.2\%), with per-cell rates from 0\% to
4.0\%. The stacked condition blocked the same 9 turns and dropped 101
memories (1.3\%). The reranker's verdict ignores the question, so it
can drop a distrusted memory again: each dropped memory was dropped
2.4 times on average. Overall, 2.1\% of stored memories were hidden at
least once, and 14.9\% of questions lost at least one retrieved
memory (8.0\% when stacked).

A benign corpus makes every drop a false alarm, but it cannot show
whether a drop hurts (Section~\ref{sec:metrics}). An external
annotator therefore labeled a sample of 50 reranker drops without
seeing answer outcomes. The annotator marked 4 drops unsure, and the
author decided them. Only 2 drops (4\%; exact 95\% CI 0.5\% to
13.7\%) were justified by content, both among those 4. Of the other
48, 26 removed a memory relevant to the question and 22 removed an
unrelated one. In 3 of the 26 relevant drops, the baseline answered
correctly and the reranker did not.

These counts explain how the reranker touched 14.9\% of questions
without a detectable accuracy cost. Most drops remove memories the
answer does not need, and a relevant drop changes the answer only
sometimes. FQR and $\Delta_d$ therefore measure different harms: the
reranker withholds evidence far more often than it changes an answer.

\subsection{Stacking (RQ3)}
\label{sec:resolution}
Stacking did not compound the cost. Stacked accuracy matches anomaly
detection's in every conversation (Figure~\ref{fig:perconv}), so the
reranker stage added no measurable change on top of the write-time
block. Its reranker stage dropped 1.3\% of judged memories against
2.2\% for the reranker alone. The two read different stores, so we do
not read this gap as an interaction. Only token cost adds up when
defenses are stacked.


\section{Discussion}
\label{sec:discussion}

\paragraph{What drives benign cost} Using an LLM did not decide how
often a defense fired. Anomaly detection and the reranker run on the
same backbone. Yet anomaly detection blocked 0.4\% of benign turns,
and the reranker dropped 2.2\% of the memories it judged, more than
five times as often. The difference lies in what each defense judges
and what it sees. Write-time screens check a whole turn for signs of
attack, which benign turns rarely show. The reranker judges whether a
short extracted memory can be trusted, without the question or the
memory's source turn.

\paragraph{Why read-time filtering misfires} Two properties amplify
the reranker's errors. It judges a memory on every retrieval, so a
wrong verdict can recur: each dropped memory was dropped 2.4 times on
average. It also judges with less context, so it distrusts benign
fragments: 48 of 50 audited drops had no grounds in their content.
Our design varies stage together with prompt and input
(Section~\ref{sec:threat-read}). We therefore report an association;
it does not show that stage alone causes the gap.

\paragraph{Accuracy hides withheld evidence} The reranker had no
detectable accuracy cost, yet it withheld memory from 14.9\% of
questions. In the audit, 26 of 50 drops removed a memory relevant to
the question, but only 3 of those turned a correct answer wrong. A
deployment judged by accuracy alone would miss this withheld
evidence. We recommend reporting FQR and token overhead alongside
accuracy and attack success rate; none of them can stand in for the
others.

\paragraph{The defense must run on the agent's model} Our earlier
version ran the reranker condition on a different model than its
baseline. That swap accounts for about two-thirds of the cost we
reported (Section~\ref{sec:result-swap}). An evaluation that runs an
LLM defense, or its condition's answers, on another model measures the
model change along with the defense. Defenses should run on the
agent's own backbone. When that is not possible, a baseline on the
defense's model separates the two effects.

\paragraph{A blocked turn changes the store} When a write-time defense
fires, it changes the memories extracted after the blocked turn. Its
accuracy delta then compares two different stores. Anomaly
detection's $+4.7$ points is such a delta. With nothing blocked,
store-to-store variation alone moved one conversation by up to 9
points. A single blocked turn can therefore look like a gain or a
loss. Evaluations of write-time defenses should report how many turns
were blocked and compare the delta with this variation.

\paragraph{What stacking does not show} Stacking did not compound cost
here, because the write-time stages blocked only three turns per
replicate. On traffic that triggers them more often, stacked filters
may interact. A benign corpus cannot test that.

\paragraph{A testable fix} The reranker's errors suggest a direct
remedy: give read-time filters the question and the memory's source
turn. This restores the context that write-time screens already have.
We did not test this, and it would raise token cost.


\section{Limitations}
\label{sec:limitations}

\paragraph{Scale} We use half of LoCoMo: five of its ten long
conversations, each truncated to 150 turns. This gives 255 questions
per replicate and 765 graded answers per condition. Interval
half-widths range from 2.4 to 4.6 points, so costs of one or two
points are hard to detect. Our nulls therefore show no detectable
cost, not zero cost. With five conversations as units, the sign-flip
test cannot give a $p$-value below 0.0625.

\paragraph{One configuration} We test one backbone, memory framework,
embedding model, and judge. Our model-swap baseline moved accuracy by
2.9 points with no defense at all, so a different backbone may change
both the costs and the FQR. The backbone is a hosted model without a
pinned version, and its extraction changed during this project
(Appendix~\ref{app:pitfalls}). Exact numbers may therefore not
reproduce.

\paragraph{Store builds} Two conversations use stores built for the
earlier version of this study, and three use stores rebuilt for this
one. Every paired $\Delta_{d,c}$ stays within one conversation and one
build period, but the pooled $\Delta_d$ averages over both. In the
rebuilt conversations, replicates were near copies, so their majority
vote adds little protection against run-to-run noise.

\paragraph{Stage and implementation} Stage varies together with prompt
and input. The anomaly detector and the reranker share a backbone but
see different inputs. Our evidence on the role of stage is therefore
associative.

\paragraph{Simplified defenses} Each defense is a minimal
implementation of a published idea. Provenance cannot fire on LoCoMo,
because its turns carry no origin information. The reranker sees
neither the question nor the source turn. Stronger implementations may
cost more or less.

\paragraph{No attacks} We measure cost only on benign traffic. We do
not measure how well any defense stops attacks, we test no act-time
defense, and we cannot study how stacked filters interact when they
fire often.

\paragraph{Measurement} The bootstrap captures run-to-run noise only
through the majority vote. One external annotator produced the hand
labels, so we report no inter-annotator agreement; the author decided
the 4 drops the annotator marked unsure. The 108 labeled adversarial
answers contain only 23 distinct texts. All 24 judge calibration items
come from conv-26, which is also an evaluation conversation. Two cells
carry documented exceptions (Appendix~\ref{app:pitfalls}). Token
overhead depends on the backbone's tokenizer and prompt lengths, and
we report no dollar cost.


\section{Conclusion}
\label{sec:conclusion}
We measured what four memory-poisoning defenses cost on benign
traffic. Every condition shared one backbone, memory pipeline, and
judge, ran three times, and passed a per-cell integrity check. No
defense showed a detectable accuracy cost, at a resolution of
$\pm$2.4 to $\pm$4.6 points. Sanitization and provenance never fired.
Anomaly detection blocked 0.4\% of turns, and its 4.7-point gain
reflects a changed store and is not a benefit. The reranker dropped
2.2\% of the memories it judged and withheld memory from 14.9\% of
questions. A hand audit found only 2 of 50 sampled drops justified.
Stacking all four defenses added only tokens, 4.5\% in total. Our
earlier version attributed a 4.4-point cost to the reranker; running
that condition on a different model caused about two-thirds of it.
Benign cost therefore depends on how it is measured. The defense must
run on the agent's model, and a write-time block must be judged
against store-to-store variation. Attack success rate shows none of
this, so benign cost belongs beside it in every defense evaluation.

\appendix

\section{Retention Under Truncation}
\label{app:retention}
Table~\ref{tab:retention} gives each conversation's full length and
question count. It then shows how many questions keep their evidence
at 150 turns, and how many remain after the cap of 75.

\begin{table}[H]
\caption{Question retention at 150-turn truncation. Turns and
Questions are for the full conversation; Evidence kept counts
questions whose evidence turns all fall within the first 150 turns.}
\label{tab:retention}
\centering
\small
\begin{tabular}{@{}lrrrr@{}}
\toprule
\textbf{Conversation} & \textbf{Turns} & \textbf{Questions}
  & \textbf{Evidence kept} & \textbf{Used (core / adv.)} \\
\midrule
conv-26 & 419 & 199 & 86 & 75 (66 / 9)   \\
conv-30 & 369 & 105 & 51 & 51 (38 / 13)  \\
conv-41 & 663 & 193 & 32 & 32 (25 / 7)   \\
conv-42 & 629 & 260 & 42 & 42 (33 / 9)   \\
conv-43 & 680 & 242 & 55 & 55 (38 / 17)  \\
\midrule
Total   & 2{,}760 & 999 & 266 & 255 (200 / 55) \\
\bottomrule
\end{tabular}
\end{table}

\section{Per-Cell Results}
\label{app:percell}
\begin{figure}[H]
\centering
\includegraphics[width=5.0in]{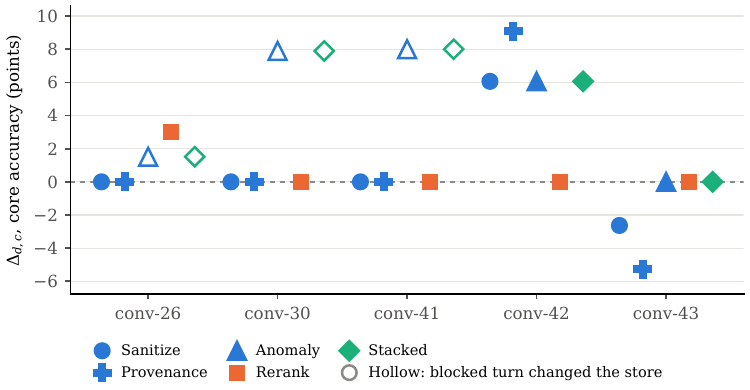}
\caption{Paired change in core accuracy per conversation, by majority
vote over three replicates. Hollow markers show conversations where
the anomaly screen blocked a turn. With nothing blocked, sanitize and
provenance still move a single conversation by up to 9 points.}
\label{fig:perconv}
\end{figure}

Table~\ref{tab:percell} lists all 105 cells we ran for this version:
the 90 main cells and the 15 new cells of the model-swap baseline. The
baseline's other 30 cells come from our earlier version, which lists
them. Figure~\ref{fig:perconv} shows the paired change per
conversation.
{\footnotesize
\setlength{\tabcolsep}{5pt}
\begin{longtable}{@{}l l r r r r r r r@{}}
\caption{Per-cell results, all 105 runs. Core is core accuracy within
the run; Adv.\ is adversarial abstention with hand labels applied.
Quar.\ counts quarantines, and FQR divides them by the decisions made
in that cell (memories judged, for the reranker stage). For the
stacked condition, both columns cover its reranker stage; its
write-time stage also blocked one turn in every conv-26, conv-30, and
conv-41 cell. Baseline-mini rows are the new cells of the model-swap
baseline.}\label{tab:percell}\\
\toprule
\textbf{Condition} & \textbf{Conv.} & \textbf{Rep.} & \textbf{Mem.} & $\boldsymbol{n_q}$ & \textbf{Core} & \textbf{Adv.} & \textbf{Quar.} & \textbf{FQR} \\
\midrule
\endfirsthead
\toprule
\textbf{Condition} & \textbf{Conv.} & \textbf{Rep.} & \textbf{Mem.} & $\boldsymbol{n_q}$ & \textbf{Core} & \textbf{Adv.} & \textbf{Quar.} & \textbf{FQR} \\
\midrule
\endhead
\bottomrule
\endlastfoot
None & conv-26 & 1 & 253 & 75 & 0.606 & 1.000 & 0 & -- \\
None & conv-26 & 2 & 253 & 75 & 0.606 & 1.000 & 0 & -- \\
None & conv-26 & 3 & 253 & 75 & 0.606 & 1.000 & 0 & -- \\
None & conv-30 & 1 & 225 & 51 & 0.579 & 0.846 & 0 & -- \\
None & conv-30 & 2 & 225 & 51 & 0.579 & 0.846 & 0 & -- \\
None & conv-30 & 3 & 225 & 51 & 0.579 & 0.846 & 0 & -- \\
None & conv-41 & 1 & 162 & 32 & 0.800 & 1.000 & 0 & -- \\
None & conv-41 & 2 & 162 & 32 & 0.800 & 1.000 & 0 & -- \\
None & conv-41 & 3 & 162 & 32 & 0.800 & 1.000 & 0 & -- \\
None & conv-42 & 1 & 247 & 42 & 0.485 & 1.000 & 0 & -- \\
None & conv-42 & 2 & 248 & 42 & 0.515 & 1.000 & 0 & -- \\
None & conv-42 & 3 & 247 & 42 & 0.485 & 0.889 & 0 & -- \\
None & conv-43 & 1 & 232 & 55 & 0.632 & 1.000 & 0 & -- \\
None & conv-43 & 2 & 253 & 55 & 0.632 & 0.941 & 0 & -- \\
None & conv-43 & 3 & 246 & 55 & 0.605 & 1.000 & 0 & -- \\
\midrule
Sanitize & conv-26 & 1 & 253 & 75 & 0.606 & 1.000 & 0 & 0.0\% \\
Sanitize & conv-26 & 2 & 253 & 75 & 0.606 & 1.000 & 0 & 0.0\% \\
Sanitize & conv-26 & 3 & 253 & 75 & 0.606 & 1.000 & 0 & 0.0\% \\
Sanitize & conv-30 & 1 & 225 & 51 & 0.579 & 0.846 & 0 & 0.0\% \\
Sanitize & conv-30 & 2 & 225 & 51 & 0.579 & 0.846 & 0 & 0.0\% \\
Sanitize & conv-30 & 3 & 225 & 51 & 0.579 & 0.846 & 0 & 0.0\% \\
Sanitize & conv-41 & 1 & 162 & 32 & 0.800 & 1.000 & 0 & 0.0\% \\
Sanitize & conv-41 & 2 & 162 & 32 & 0.800 & 1.000 & 0 & 0.0\% \\
Sanitize & conv-41 & 3 & 172 & 32 & 0.720 & 1.000 & 0 & 0.0\% \\
Sanitize & conv-42 & 1 & 265 & 42 & 0.545 & 1.000 & 0 & 0.0\% \\
Sanitize & conv-42 & 2 & 258 & 42 & 0.545 & 1.000 & 0 & 0.0\% \\
Sanitize & conv-42 & 3 & 280 & 42 & 0.606 & 0.889 & 0 & 0.0\% \\
Sanitize & conv-43 & 1 & 230 & 55 & 0.526 & 0.941 & 0 & 0.0\% \\
Sanitize & conv-43 & 2 & 242 & 55 & 0.579 & 0.941 & 0 & 0.0\% \\
Sanitize & conv-43 & 3 & 247 & 55 & 0.605 & 0.941 & 0 & 0.0\% \\
\midrule
Provenance & conv-26 & 1 & 253 & 75 & 0.606 & 1.000 & 0 & 0.0\% \\
Provenance & conv-26 & 2 & 253 & 75 & 0.606 & 1.000 & 0 & 0.0\% \\
Provenance & conv-26 & 3 & 253 & 75 & 0.606 & 1.000 & 0 & 0.0\% \\
Provenance & conv-30 & 1 & 225 & 51 & 0.579 & 0.846 & 0 & 0.0\% \\
Provenance & conv-30 & 2 & 225 & 51 & 0.579 & 0.846 & 0 & 0.0\% \\
Provenance & conv-30 & 3 & 225 & 51 & 0.579 & 0.846 & 0 & 0.0\% \\
Provenance & conv-41 & 1 & 162 & 32 & 0.800 & 1.000 & 0 & 0.0\% \\
Provenance & conv-41 & 2 & 162 & 32 & 0.800 & 1.000 & 0 & 0.0\% \\
Provenance & conv-41 & 3 & 162 & 32 & 0.800 & 1.000 & 0 & 0.0\% \\
Provenance & conv-42 & 1 & 241 & 42 & 0.606 & 0.889 & 0 & 0.0\% \\
Provenance & conv-42 & 2 & 230 & 42 & 0.545 & 1.000 & 0 & 0.0\% \\
Provenance & conv-42 & 3 & 253 & 42 & 0.545 & 1.000 & 0 & 0.0\% \\
Provenance & conv-43 & 1 & 232 & 55 & 0.579 & 0.941 & 0 & 0.0\% \\
Provenance & conv-43 & 2 & 245 & 55 & 0.579 & 1.000 & 0 & 0.0\% \\
Provenance & conv-43 & 3 & 219 & 55 & 0.579 & 1.000 & 0 & 0.0\% \\
\midrule
Anomaly & conv-26 & 1 & 268 & 75 & 0.621 & 1.000 & 1 & 0.7\% \\
Anomaly & conv-26 & 2 & 268 & 75 & 0.621 & 1.000 & 1 & 0.7\% \\
Anomaly & conv-26 & 3 & 268 & 75 & 0.621 & 1.000 & 1 & 0.7\% \\
Anomaly & conv-30 & 1 & 210 & 51 & 0.658 & 0.923 & 1 & 0.7\% \\
Anomaly & conv-30 & 2 & 215 & 51 & 0.658 & 1.000 & 1 & 0.7\% \\
Anomaly & conv-30 & 3 & 216 & 51 & 0.658 & 1.000 & 1 & 0.7\% \\
Anomaly & conv-41 & 1 & 164 & 32 & 0.880 & 1.000 & 1 & 0.7\% \\
Anomaly & conv-41 & 2 & 164 & 32 & 0.880 & 1.000 & 1 & 0.7\% \\
Anomaly & conv-41 & 3 & 164 & 32 & 0.880 & 1.000 & 1 & 0.7\% \\
Anomaly & conv-42 & 1 & 254 & 42 & 0.545 & 1.000 & 0 & 0.0\% \\
Anomaly & conv-42 & 2 & 259 & 42 & 0.545 & 1.000 & 0 & 0.0\% \\
Anomaly & conv-42 & 3 & 227 & 42 & 0.545 & 1.000 & 0 & 0.0\% \\
Anomaly & conv-43 & 1 & 230 & 55 & 0.579 & 0.941 & 0 & 0.0\% \\
Anomaly & conv-43 & 2 & 237 & 55 & 0.605 & 1.000 & 0 & 0.0\% \\
Anomaly & conv-43 & 3 & 221 & 55 & 0.579 & 0.941 & 0 & 0.0\% \\
\midrule
Rerank & conv-26 & 1 & 253 & 75 & 0.636 & 1.000 & 24 & 3.2\% \\
Rerank & conv-26 & 2 & 263 & 75 & 0.636 & 1.000 & 17 & 2.3\% \\
Rerank & conv-26 & 3 & 253 & 75 & 0.636 & 1.000 & 24 & 3.2\% \\
Rerank & conv-30 & 1 & 225 & 51 & 0.579 & 0.846 & 10 & 2.0\% \\
Rerank & conv-30 & 2 & 225 & 51 & 0.579 & 0.846 & 10 & 2.0\% \\
Rerank & conv-30 & 3 & 225 & 51 & 0.579 & 0.846 & 10 & 2.0\% \\
Rerank & conv-41 & 1 & 162 & 32 & 0.800 & 1.000 & 3 & 0.9\% \\
Rerank & conv-41 & 2 & 162 & 32 & 0.800 & 1.000 & 3 & 0.9\% \\
Rerank & conv-41 & 3 & 162 & 32 & 0.800 & 1.000 & 3 & 0.9\% \\
Rerank & conv-42 & 1 & 241 & 42 & 0.455 & 1.000 & 4 & 1.0\% \\
Rerank & conv-42 & 2 & 246 & 42 & 0.576 & 1.000 & 0 & 0.0\% \\
Rerank & conv-42 & 3 & 245 & 42 & 0.485 & 1.000 & 0 & 0.0\% \\
Rerank & conv-43 & 1 & 252 & 55 & 0.605 & 1.000 & 22 & 4.0\% \\
Rerank & conv-43 & 2 & 238 & 55 & 0.605 & 0.941 & 20 & 3.6\% \\
Rerank & conv-43 & 3 & 236 & 55 & 0.579 & 0.941 & 19 & 3.5\% \\
\midrule
Stacked & conv-26 & 1 & 268 & 75 & 0.621 & 1.000 & 3 & 0.4\% \\
Stacked & conv-26 & 2 & 268 & 75 & 0.621 & 1.000 & 3 & 0.4\% \\
Stacked & conv-26 & 3 & 268 & 75 & 0.621 & 1.000 & 3 & 0.4\% \\
Stacked & conv-30 & 1 & 216 & 51 & 0.658 & 1.000 & 7 & 1.4\% \\
Stacked & conv-30 & 2 & 216 & 51 & 0.658 & 1.000 & 7 & 1.4\% \\
Stacked & conv-30 & 3 & 211 & 51 & 0.632 & 1.000 & 7 & 1.4\% \\
Stacked & conv-41 & 1 & 164 & 32 & 0.880 & 1.000 & 0 & 0.0\% \\
Stacked & conv-41 & 2 & 164 & 32 & 0.880 & 1.000 & 0 & 0.0\% \\
Stacked & conv-41 & 3 & 164 & 32 & 0.880 & 1.000 & 0 & 0.0\% \\
Stacked & conv-42 & 1 & 260 & 42 & 0.424 & 1.000 & 1 & 0.2\% \\
Stacked & conv-42 & 2 & 249 & 42 & 0.515 & 1.000 & 0 & 0.0\% \\
Stacked & conv-42 & 3 & 251 & 42 & 0.576 & 1.000 & 1 & 0.2\% \\
Stacked & conv-43 & 1 & 240 & 55 & 0.605 & 1.000 & 21 & 3.8\% \\
Stacked & conv-43 & 2 & 249 & 55 & 0.605 & 1.000 & 22 & 4.0\% \\
Stacked & conv-43 & 3 & 239 & 55 & 0.579 & 0.941 & 26 & 4.7\% \\
\midrule
Baseline-mini & conv-26 & 1 & 309 & 75 & 0.652 & 1.000 & 0 & -- \\
Baseline-mini & conv-26 & 2 & 292 & 75 & 0.652 & 1.000 & 0 & -- \\
Baseline-mini & conv-26 & 3 & 316 & 75 & 0.636 & 1.000 & 0 & -- \\
Baseline-mini & conv-30 & 1 & 282 & 51 & 0.658 & 0.923 & 0 & -- \\
Baseline-mini & conv-30 & 2 & 248 & 51 & 0.658 & 1.000 & 0 & -- \\
Baseline-mini & conv-30 & 3 & 253 & 51 & 0.658 & 1.000 & 0 & -- \\
Baseline-mini & conv-41 & 1 & 250 & 32 & 0.800 & 1.000 & 0 & -- \\
Baseline-mini & conv-41 & 2 & 234 & 32 & 0.800 & 1.000 & 0 & -- \\
Baseline-mini & conv-41 & 3 & 243 & 32 & 0.760 & 0.857 & 0 & -- \\
Baseline-mini & conv-42 & 1 & 247 & 42 & 0.485 & 1.000 & 0 & -- \\
Baseline-mini & conv-42 & 2 & 248 & 42 & 0.455 & 1.000 & 0 & -- \\
Baseline-mini & conv-42 & 3 & 247 & 42 & 0.485 & 1.000 & 0 & -- \\
Baseline-mini & conv-43 & 1 & 232 & 55 & 0.500 & 0.941 & 0 & -- \\
Baseline-mini & conv-43 & 2 & 253 & 55 & 0.579 & 1.000 & 0 & -- \\
Baseline-mini & conv-43 & 3 & 246 & 55 & 0.553 & 1.000 & 0 & -- \\
\end{longtable}
}

\section{Implementation Pitfalls}
\label{app:pitfalls}
We report six problems. Five returned plausible numbers instead of
errors. The sixth crashed, but a silent version would have passed
unnoticed.

\paragraph{Defenses on a different model}
Our earlier version stated that both LLM defenses ran on the agent's
backbone. In fact, the anomaly and reranker verdicts, and the answers
in the reranker and stacked conditions, came from
\texttt{gpt-4o-mini}. No call failed, and every number looked
plausible. The swap alone produced most of the reranker cost we then
reported (Section~\ref{sec:result-swap}). A live check on every
proxy-log line now stops any run whose call reaches a model other than
the one assigned to its step.

\paragraph{The store depends on the date and the model}
Mem0 2.0.15 writes the current UTC date into every extraction prompt.
Two stores of one conversation built on different days agreed on at
most 47 of 150 turns. Stores built on the same day agreed on at least
their first 53 turns, and most on 144 to 150. The hosted backbone also
changed during the project. Stores from our earlier version differed
even when built on the same day. Rerunning our earlier code later on
one conversation stored 225 memories instead of the original 282,
close to the 213 from our current code, although every argument
passed to Mem0 was identical. We attribute this change to the hosted
model. We therefore fixed the date for all rebuilt stores and kept
each conversation within one build period
(Section~\ref{sec:protocol}).

\paragraph{Failed calls that fail open}
Our earlier code read a verdict that still failed after retries as
\texttt{BENIGN} or \texttt{TRUST}, so a defense could fail open
without notice. Interrupted runs could also lose turns, or carry an
earlier attempt's last messages into a rerun. The current code stops
the run on a failed call. Only our harness retries, and it counts
every retry. The integrity check catches lost turns and leftover
history (Section~\ref{sec:integrity}). Two cells keep documented
exceptions. In one cell of the model-swap baseline, taken from the
earlier version, 2 of 320 reranker verdicts defaulted to
\texttt{TRUST} after 149 rate-limit and server errors. In one main
cell, the proxy retried an abandoned call 79 minutes after the cell
finished, and the cell still recorded exactly 750 decisions.

\paragraph{Silent failure on a reasoning model}
Our first backbone was a larger Gemini reasoning model, and extraction
used Mem0's default limit of 2,000 output tokens. On conv-26 and
conv-30, the baseline stored 82 and 67 memories. After we switched to
the final backbone and set a 4,000-token limit, the same conversations
stored 309 and 282. Only 4 of the 300 extraction calls came within 50
tokens of the limit, so truncation explains little of this gap. The
rest reflects the model change or extraction output that Mem0
discarded. We did not log extraction outputs, so we cannot separate
the two. The pipeline raised no error either way.

\paragraph{A narrow detector taken for ground truth}
Our first abstention scorer matched a short list of refusal phrases.
Of the 52 answers it scored as failures that the annotator labeled, 17
were correct refusals phrased in ways it missed. Reading the failing
answers exposed the error. We now hand-label the answers the regex
scores as failures (Section~\ref{sec:judge}).

\paragraph{Defense composition on a wrapped store}
In the stacked condition, each defense wraps the next one, not the
store. Our reranker assumed direct access to the store and raised an
error on every question. It now walks the wrapper chain to reach the
store. We fixed this before the reported runs.


\section*{References}

{
\small

[1] Dong, S., Xu, S., He, P., Li, Y., Tang, J., Liu, T., Liu, H.\ \& Xiang, Z.\ (2025) A practical memory injection attack against LLM agents. {\it arXiv preprint arXiv:2503.03704}.

[2] Srivastava, S.S.\ \& He, H.\ (2025) MemoryGraft: Persistent compromise of LLM agents via poisoned experience retrieval. {\it arXiv preprint arXiv:2512.16962}.

[3] Chen, Z., Xiang, Z., Xiao, C., Song, D.\ \& Li, B.\ (2024) AgentPoison: Red-teaming LLM agents via poisoning memory or knowledge bases. In {\it Advances in Neural Information Processing Systems 37}, pp.\ 130185--130213.

[4] Zou, W., Geng, R., Wang, B.\ \& Jia, J.\ (2025) PoisonedRAG: Knowledge corruption attacks to retrieval-augmented generation of large language models. In {\it Proceedings of the 34th USENIX Security Symposium}, pp.\ 3827--3844. Seattle, WA.

[5] Chu, K.\ (2026) A systematic survey of security threats and defenses in LLM-based AI agents: A layered attack surface framework. {\it arXiv preprint arXiv:2604.23338}.

[6] Zhang, H., Huang, J., Mei, K., Yao, Y., Wang, Z., Zhan, C., Wang, H.\ \& Zhang, Y.\ (2025) Agent Security Bench (ASB): Formalizing and benchmarking attacks and defenses in LLM-based agents. In {\it International Conference on Learning Representations}.

[7] Dash, P., Ge, T., Jain, A., Shah, T.\ \& Shang, Z.\ (2026) From untrusted input to trusted memory: A systematic study of memory poisoning attacks in LLM agents. {\it arXiv preprint arXiv:2606.04329}.

[8] Debenedetti, E., Shumailov, I., Fan, T., Hayes, J., Carlini, N., Fabian, D., Kern, C., Shi, C., Terzis, A.\ \& Tram\`er, F.\ (2025) Defeating prompt injections by design. {\it arXiv preprint arXiv:2503.18813}.

[9] Sunil, B.D., Sinha, I., Maheshwari, P., Todmal, S., Mallik, S.\ \& Mishra, S.\ (2026) Memory poisoning attack and defense on memory based LLM-agents. {\it arXiv preprint arXiv:2601.05504}.

[10] Zhou, H., Lee, K.-H., Zhan, Z., Chen, Y., Li, Z., Wang, Z., Haddadi, H.\ \& Yilmaz, E.\ (2025) TrustRAG: Enhancing robustness and trustworthiness in retrieval-augmented generation. {\it arXiv preprint arXiv:2501.00879}.

[11] Sharma, T.\ (2026) SMSR: Certified defence against runtime memory poisoning in persistent LLM agent systems. {\it arXiv preprint arXiv:2606.12703}.

[12] Gowda, I.\ (2026) MEMSAD: Gradient-coupled anomaly detection for memory poisoning in retrieval-augmented agents. {\it arXiv preprint arXiv:2605.03482}.

[13] Xiang, C., Wu, T., Zhong, Z., Wagner, D., Chen, D.\ \& Mittal, P.\ (2024) Certifiably robust RAG against retrieval corruption. {\it arXiv preprint arXiv:2405.15556}.

[14] Wei, Q., Yang, T., Wang, Y., Li, X., Li, L., Yin, Z., Zhan, Y., Holz, T., Lin, Z.\ \& Wang, X.\ (2025) A-MemGuard: A proactive defense framework for LLM-based agent memory. {\it arXiv preprint arXiv:2510.02373}.

[15] Beurer-Kellner, L., Buesser, B., Cre\c{t}u, A.-M., Debenedetti, E., Dobos, D., Fabian, D., Fischer, M., Froelicher, D., Grosse, K., Naeff, D., Ozoani, E., Paverd, A., Tram\`er, F.\ \& Volhejn, V.\ (2025) Design patterns for securing LLM agents against prompt injections. {\it arXiv preprint arXiv:2506.08837}.

[16] Chhikara, P., Khant, D., Aryan, S., Singh, T.\ \& Yadav, D.\ (2025) Mem0: Building production-ready AI agents with scalable long-term memory. {\it arXiv preprint arXiv:2504.19413}.

[17] Wu, D., Wang, H., Yu, W., Zhang, Y., Chang, K.-W.\ \& Yu, D.\ (2025) LongMemEval: Benchmarking chat assistants on long-term interactive memory. In {\it International Conference on Learning Representations}, pp.\ 86809--86836.

[18] Cherif, A.\ (2026) AgentMemBench: A systematic benchmark for evaluating long-term memory management strategies in conversational AI agents. {\it arXiv preprint arXiv:2608.00009}.

[19] Wolff, B.\ \& Bennati, J.\ (2026) Cost and accuracy of long-term memory in distributed multi-agent systems based on large language models. {\it arXiv preprint arXiv:2601.07978}.

[20] Maharana, A., Lee, D.-H., Tulyakov, S., Bansal, M., Barbieri, F.\ \& Fang, Y.\ (2024) Evaluating very long-term conversational memory of LLM agents. In {\it Proceedings of the 62nd Annual Meeting of the Association for Computational Linguistics}, pp.\ 13851--13870.

[21] Efron, B.\ \& Tibshirani, R.J.\ (1993) {\it An Introduction to the Bootstrap}. New York, NY: Chapman \& Hall/CRC.

[22] McNemar, Q.\ (1947) Note on the sampling error of the difference between correlated proportions or percentages. {\it Psychometrika} {\bf 12}(2):153--157.

[23] Debenedetti, E., Zhang, J., Balunovi\'c, M., Beurer-Kellner, L., Fischer, M.\ \& Tram\`er, F.\ (2024) AgentDojo: A dynamic environment to evaluate prompt injection attacks and defenses for LLM agents. In {\it Advances in Neural Information Processing Systems 37, Datasets and Benchmarks Track}.




\end{document}